# Data Accuracy Estimation for Spatially Correlated Data in Wireless Sensor Networks under Distributed Clustering

Jyotirmoy Karjee , H.S Jamadagni
Centre for Electronics Design and Technology
Indian Institute of Science
Bangalore, India
kjyotirmoy@cedt.iisc.ernet.in, hsjam@cedt.iisc.ernet.in

*Abstract—Objective*-The main purpose of this paper is to construct a distributed clustering algorithm such that each distributed cluster can perform the data accuracy at their respective cluster head node before data aggregation and transmit the data to the sink node. *Design approach/Procedure* – We investigate that the data are spatially correlated among the sensor nodes which form the clusters in the spatial domain. Due to high correlation of data, these clusters of sensor nodes are overlapped in the spatial domain. To overcome this problem, we construct a distributed clustering algorithm with non-overlapping irregular clusters in the spatial domain. Then each distributed cluster can perform data accuracy at the cluster head node and finally send the data to the sink node. *Findings-* Simulation result shows the associate sensor nodes of each distributed cluster and clarifies their data accuracy profile in the spatial domain. We demonstrate the simulation results for a single cluster to verify that their exist an optimal cluster which give approximately the same data accuracy level achieve by the single cluster. Moreover we find that as the distance from the tracing point to the number of sensor node increases the data accuracy decreases. *Design Limitations* – This model is only applicable to estimate data accuracy for distributed clusters where the sensed data are assumed to be spatially correlated with approximately same variations. *Practical implementation* – Measure the moisture content in the distributed agricultural field. *Inventive/Novel idea-* This is the first time that a data accuracy model is performed for the distributed clusters before data aggregation at the cluster head node which can reduce data redundancy and communication overhead.

I. INTRODUCTION

Wireless sensor network has made a drastic change in communications for the last several years. One of the vital tasks of wireless sensor network is to sense or measure the physical phenomenon of data such as measurement of humidity, temperature, seismic event etc from the environment [1]. Physical phenomenon of data is measured or sense by a device called sensor nodes which are capable to sense, process and communicate the data through out the network. Since most of the data are spatially correlated [2] among them, the sensor nodes form clusters in the sensor field to reduce data collection cost [3]. According to literature survey, LEACH [4] gives a clear idea about how dynamically cluster and cluster head are created according to a priori probability. Finally cluster head aggregate all the data and send it to the sink node. Similarly SEP [5] demonstrates the formation of cluster in heterogeneous sensor networks. Since data correlation in wireless sensor networks shows Gaussian distribution with zero mean, literature [6] shows the spatial correlation among data in high in sensor networks but it lags the practical implementation of analyzing the correlated data for transmitting the packets for communication. Literature [7] proposes a grid based spatial correlation clustering method where the entire cluster is equipped in a grid sensor field. However this type of model rarely happens in an original scenario in wireless sensor networks. Moreover literature [8] proposes a disk-shaped circular cluster where sensor nodes are grouped into disjoint sets each managed by a designated cluster head which lags the practical shape of a cluster. As most cases the cluster formation are irregular in shape in the spatial domain. Hence in this paper we propose a foundation of distributed clustering algorithm which is much more practical than the previous work done in the spatial domain. In our model, we propose a spatially correlated distributed irregular non overlapping cluster formation in the spatial domain. These distributed irregular cluster formation in the spatial domain is much more practical model in original scenario than the previous literature discussed above.

Most of the work done till today is based upon the fact that the sink node or the base station is responsible for estimating the data accuracy for physically sensed data by sensor nodes [9, 10, 11] .Therefore it is applicable for one hop communication where the raw data are sensed and measured by the sensor nodes and directly transmitted to the sink node. Again we propose a model [12] for data accuracy where we have considered two hop communications in which physical phenomenon of sensed data is transmitted via intermediate node called cluster head (*CH*)[18]node. But in this paper we propose a distributed clustering algorithm where each cluster can perform data accuracy at their respective *CH* node and finally send the data to the sink node. Each distributed cluster is responsible for sensing and measuring the physical phenomenon of data in the sensor region.

The main goal of this paper is to estimate data accuracy for each distributed cluster before data aggregation [19] at their respective *CH* node which can reduce the data redundancy and communication overhead. However to the best understanding of the authors, there is no work done so far on verifying the data accuracy for distributed cluster before data aggregation [21, 22] at their respective *CH* node. Since from the literature survey we have seen that most of the work done till today is that data from cluster of sensor nodes directly send to *CH* node for aggregation without verifying its accuracy. Hence it is important that the most precise or accurate data send by the distributed cluster can aggregate at their respective C*H* node before transmitting to the sink node and not aggregating all the redundant data at *CH* node. The data send by each distributed cluster should first verify its accuracy level at their respective *CH* node then only the data get aggregates and finally send to the sink node. Since *CH* node verifies the data accuracy for their respective distributed cluster, it may reduce the power consumption and increase the lifetime of the networks.

Another important reason for estimating data accuracy for each distributed cluster before data aggregation at their respective *CH* node, if some of the sensor nodes in the distributed cluster get malicious [20]. If some of the sensor nodes become malicious in the distributed cluster, then it can sense and read inaccurate data. These inaccurate data send by the malicious nodes gets aggregated with the other correct data results in inaccurate (incorrect) data aggregation at the *CH* node of their respective cluster and finally send to the sink node. This may increase the power consumption, data redundancy and communication overhead in the distributed network. It results very high or low variations of the estimated data accuracy value compare to the actual variations of estimated data accuracy value at the *CH* node. Hence to overcome this problem, it is important to estimate the data accuracy at *CH* node for distributed cluster before data aggregation and send the accurate data to the sink node. In our model we assume that the sensed data are spatially correlated with approximately the same variations in each distributed cluster and the sensor nodes are appropriate to sense the correct data. We verify estimated data accuracy with approximately same variations at the *CH* node for each distributed cluster.

In our model, each distributed cluster is responsible to sense the physical phenomenon of data such as moisture content of soil in the sensor region. Once the data accuracy is processed by *CH* node for each distributed cluster, it transmits the estimated accurate data to the sink node. From the literature survey, it is clear that only the sensor nodes are responsible to sense the physical phenomenon of data and not the sink node. But in our model not only sensor nodes are responsible to sense the physical phenomenon of data but the *CH* node can also do the sensing phenomenon in each distributed cluster. We investigate how each distributed cluster can sense the physical phenomenon of data to estimate the data accuracy in the sensor field. Literature [9, 13] has given some approaches regarding jointly sensing nodes which gives an idea about how the raw data is sensed by the jointly sensing nodes and how the number of jointly sensing nodes affects the data accuracy. However they address this problem if only sensing nodes are responsible to retrieve physical phenomenon of data where they investigate to find a proper number and positions of jointly sensing nodes. But in our model, we consider both the sensor nodes and the *CH* node which forms each distributed cluster in the sensor field are sensing the physical phenomenon such as humidity or moisture content of the soil. Since we verify data accuracy for each distributed cluster in the sensor field, there exit an optimal cluster which gives approximately the same data accuracy level achieve by each cluster.

Rest of the paper is given as follows. In section II, we construct a data correlation model for sensor nodes in spatial domain. These data correlation can give rise to overlapping of clusters in the sensor region. Hence to overcome this problem, we propose a distributed clustering algorithm with non overlapping irregular clusters in the spatial domain. Then we perform data accuracy for each distributed cluster at *CH* node before data aggregation in the sensor region. In section III, we verify simulation results for distributed clusters. We demonstrate results how each distributed cluster are formed with their respective associate nodes and their data accuracy. Then we show the performance model of a single cluster with respect to data accuracy. Finally we conclude our work in section IV.

## II. SYSTEM MODEL

In this section, sensor nodes deployment strategies are done where the sensor nodes form distributed clusters which are capable to perform data accuracy in the spatial domain. We propose an algorithm for distributed clusters which perform data accuracy at the cluster head node where the data are spatially correlated and finally send the data to the sink node. Let a set of sensor nodes are deterministically deployed uniformly over a sensor region Z. These set of sensor nodes forms the cluster head nodes [18] for the distributed clusters equipped with additional energy resource [5]. Since *CH* node perform the data accuracy for the respective distributed clusters, we set the *CH* node with additional energy resource and distributed deterministically in the sensor field. Again another set of sensor nodes are randomly deployed over the sensor region Z and are called normal nodes [5]. Normal nodes form the distributed cluster along with their respective *CH* node which can sense and measure the spatially correlated data and estimate the data accuracy at the *CH* node.*CH* node has more energy resource than the normal nodes because *CH* nodes has to estimate the data accuracy for the cluster. Thus *CH* nodes and normal nodes form the total set of sensor nodes represented as *L* with $Z \subseteq R^2$ where ||*L*|| can be represented as total number of sensor nodes. They are capable for sensing and measuring the spatially correlated data in the sensor region Z. For example, we measure the moisture content of soil at different locations of sensor region Z. Generally there are much more variations in measurement of moisture content at different locations in the sensor field. Some places the water (or moisture) content in the soil are more than other different places where the water (moisture) content is less. Thus there are variations of monitoring the measurement of moisture content in the soil at different places in the sensor region Z.

*A.   Data Correlation for sensor nodes in Spatial Domain*

We consider reference values for higher concentration of moisture content at different places of sensor region Z. Suppose the reference values are called tracing points [20] and can be represented as $S^i$ where $i=1, 2, 3...n$ are the number of tracing points at different locations in the sensor field with higher variations .The tracing points can be located at the different places of sensor field where the moisture content is high. For example, water (or moisture) content in the soil can be higher at different locations of the sensor field. It is considered as reference values for tracing points at different locations in sensor the region Z. Although the data are spatially correlated in the sensor region, there are variations in measurement for concentratation of data (moisture content) at different places in sensor region Z. The higher concentratation of data has higher variations with respect to lower variations of data at different places. In spatial domain, data correlation depends upon the distance between the tracing points to the sensor nodes and the distance between jointly sensing nodes [13]. Thus we have two points to note in our work. Firstly, data correlation decreases as the distance between the tracing points (or reference values) to the sensor nodes increases. Secondly, data correlation decreases as the distance between jointly sensing nodes increases. Thus data correlation is more when the sensor nodes are close to each other.

Since these tracing points has higher concentratation of moisture content with higher variations, the sensor nodes can sense the higher variation of tracing points (or reference values )at different locations in sensor field . There may be higher or lower variations of data (moisture) measurement in spatial domain where the data are spatially correlated in the sensor field. Thus if the distance from the tracing point to the sensor nodes increases, the variations of the data correlation also get decreases.

We represent a single tracing point where $S^i$ for $i=1$ sensed by the sensor nodes $S_i$ and $S_j$ where they sense and do measurement over a window frame of time $T$ to capture the continuous data sample with $S_i=\{ s_{i1},\ s_{i2},\ s_{i3},\ ........s_{in} \}$ and $S_j=\{s_{j1},\ s_{j2},\ s_{j3},\ ........s_{jn}\}$ respectively. The data correlation is strong when the tracing point is sensed by the sensor nodes $S_i$ and $S_j$ located near to each other. The data correlation decreases as sensor node $S_i$ and $S_j$ are far apart from tracing point. We compute the mean of the sampled data of sensor nodes as follows

$$\bar{S}_i = \frac{1}{n}\sum_{k=1}^{n} s_{ik} \text{ and } \bar{S}_j = \frac{1}{n}\sum_{k=1}^{n} s_{jk}$$

Variance of the sample data collected by nodes $S_i$ and $S_j$ can be given as

$$\text{var}(S_i) = \frac{1}{n-1}\sum_{k=1}^{n}(s_{ik}-\bar{S}_i)^2 \quad (1)$$

and $$\text{var}(S_j) = \frac{1}{n-1}\sum_{k=1}^{n}(s_{jk}-\bar{S}_j)^2 \quad (2)$$

The covariance is given as

$$\text{cov}(S_i, S_j) = \frac{1}{(n-1)}\sum_{k=1}^{n}(s_{ik}-\bar{S}_i)(s_{jk}-\bar{S}_j) \quad (3)$$

The correlation coefficient ($\rho_{S_i,S_j}$) for correlation between data sensed by the sensor nodes $S_i$ and $S_j$ for the tracing points can be given by

$$\rho_{S_i,S_j} = \frac{\text{cov}(S_i, S_j)}{\text{var}(S_i).\text{var}(S_j)}$$

$$\rho_{S_i,S_j} = \frac{\frac{1}{(n-1)}\sum_{k=1}^{n}(s_{ik}-\bar{S}_i)(s_{jk}-\bar{S}_j)}{\left[\frac{1}{n-1}\sum_{k=1}^{n}(s_{ik}-\bar{S}_i)^2\right]\left[\frac{1}{n-1}\sum_{k=1}^{n}(s_{jk}-\bar{S}_j)^2\right]} \quad (4)$$

The *equation-no 4* shows the data correlation coefficient for nodes $S_i$ and $S_j$ in the spatial domain. Similarly from the co-variance model [16], we get the correlation coefficient ($\rho_{S_i,S_j}$) for the data in spatial domain.

$$E[S_i, S_j] = \text{cov}[S_i, S_j] = \sigma_{S^i}^2 \text{corr}[S_i, S_j] = \sigma_{S^i}^2.\rho[S_i, S_j]$$

$$\rho[S_i, S_j] = \frac{\text{cov}[S_i, S_j]}{\sigma_{S^i}^2} = \frac{E[S_i, S_j]}{\sigma_{S^i}^2} \quad (5)$$

Again from the power exponential model [16,17], we get the correlation coefficient function between node $S_i$ $(x_i, y_i)$ and node $S_j$ $(x_j, y_j)$ as follows

$$\rho[S_i, S_j] = e^{-\left(\frac{d}{\theta_1}\right)^{\theta_2}} \quad (6)$$

We define a threshold $\tau$ which can determine whether the data are spatially correlated among the sensor nodes to trace the higher variations of data (called as tracing points) in the spatial domain. $\theta_1$ is called a '*Range parameter*' which controls how fast the spatially correlated data decays with the distance. $\theta_2$ is called a '*Smoothness parameter*' which controls the geometrical properties of wireless sensor field.

If $\rho[S_i, S_j] \geq \tau$, Data are strongly correlated in spatial domain for nodes $S_i$ and $S_j$.

If $\rho[S_i, S_j] < \tau$, Data are weakly correlated in spatial domain for nodes $S_i$ and $S_j$.

From *equation no. (4), (5) and (6),* we can derive the correlation coefficient $\rho[S_i, S_j]$ of data for nodes $S_i$ and $S_j$ represented as follows:

$$\rho[S_i, S_j] = \frac{\text{cov}[S_i, S_j]}{\sigma_{S^i}^2} = e^{-\left(\frac{d}{\theta_1}\right)^{\theta_2}} \quad (7)$$

When the data are strongly correlated for nodes $S_i$ and $S_j$ in the spatial domain we have

$$\rho[S_i, S_j] = \frac{\text{cov}[S_i, S_j]}{\sigma_{S^i}^2} = e^{-\left(\frac{d}{\theta_1}\right)^{\theta_2}} \geq \tau \quad (8)$$

From the *equation no (8)*, we can derive as following

$$e^{-\left(\frac{d}{\theta_1}\right)^{\theta_2}} \geq \tau$$

or $\quad -\left(\frac{d}{\theta_1}\right)^{\theta_2} \geq \log(\tau)$

or $\quad \left(\frac{d}{\theta_1}\right)^{\theta_2} \leq \log\left(\frac{1}{\tau}\right)$

or $\quad d^2 \leq \theta_1^2 \sqrt[\theta_2]{\left(\log\left(\frac{1}{\tau}\right)\right)^2} \quad (9)$

where the Euclidean distance between the node $S_i (x_i, y_i)$ and node $S_j (x_j, y_j)$ as follows

$$d^2 = (x_i - x_j)^2 + (y_i - y_j)^2$$

Put the value of $d^2$ in equation no. (9), we get

$$(x_i - x_j)^2 + (y_i - y_j)^2 \leq \theta_1^2 \sqrt[\theta_2]{\left(\log\left(\frac{1}{\tau}\right)\right)^2} \quad (10)$$

Compare *equation no.* (10) with equation of circle with cluster head at the centre with the radius of the cluster $r$, we get

$$(x_i - x_j)^2 + (y_i - y_j)^2 = r^2 \quad (11)$$

From *equation no. (10) and (11)*, we get

$$r^2 \leq \theta_1^2 \sqrt[\theta_2]{\left(\log\left(\frac{1}{\tau}\right)\right)^2} \quad (12)$$

The *equation no.* (12), shows the relation between the radius of the cluster and the threshold value of spatially correlated data. The radius of the cluster depends upon the threshold value $\tau$, $\theta_1$ and $\theta_2$. If the value of threshold $\tau$ increases, the radius of the cluster from the *CH* node located at the centre of the cluster get decreases. So we have taken the appropriate value of $\theta_1$, $\theta_2$ and the threshold value $\tau$ to maintain a good correlation of data between sensor nodes for the clusters.

**B. Distributed Cluster Formation in Spatial Domain**

We consider a square field of area with $Z = Z_1 \times Z_2$ where the cluster head (*CH*) node are deterministically deployed uniformly and the normal nodes are deployed randomly in the sensor field Z which form the distributed cluster. Since the number of cluster head node deployed in the sensor region is known, we get the same number of clusters as the number of cluster head nodes. We are interested in measuring the moisture content profile in each cluster embedded in the sensor field Z. Thus we assume that every cluster has a single tracing point. Every cluster in the sensor field is responsible for sensing and measuring the physical phenomenon of data for the tracing point value. The highly correlated data among the sensor nodes and the *CH* node forms the cluster. The *CH* node located at the centre of each cluster performs the estimation of data accuracy and finally send the data to the sink node. The number of tracing points is equal to the number of cluster head nodes. Hence in our model numbers of sensor (normal) nodes are considered to be more than the number of cluster head nodes.

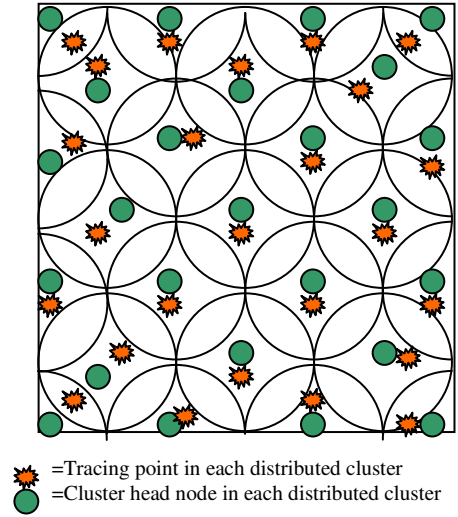

=Tracing point in each distributed cluster
=Cluster head node in each distributed cluster

*Figure 1: Overlapping clusters in sensor region*

Thus in the square sensor field Z, every cluster are embedded in the sensor field which are capable to sense their respective tracing point (to measure the high variation of correlated data) distributed uniformly as shown in

Figure-1. Thus the known number of clusters formed in the sensor region Z can be represented as $N$ as follows

$$\left(\left\lfloor\frac{Z_1}{2r}\right\rfloor\left\lfloor\frac{Z_2}{2r}\right\rfloor + \left\lfloor\frac{Z_1}{2r}+1\right\rfloor\left\lfloor\frac{Z_2}{2r}+1\right\rfloor\right) \leq \text{Number of}$$

$$\text{Clusters} \leq \left(\left\lceil\frac{Z_1}{2r}\right\rceil\left\lceil\frac{Z_2}{2r}\right\rceil + \left\lceil\frac{Z_1}{2r}+1\right\rceil\left\lceil\frac{Z_2}{2r}+1\right\rceil\right)$$

Since the sensor field is square, $Z_1 = Z_2 = W$

$$\left\lfloor \frac{W^2}{2\left(\theta_1^2 \sqrt[\theta_2]{\left(\log\left(\frac{1}{\tau}\right)\right)^2}\right)} + \frac{W}{\theta_1 \sqrt[\theta_2]{\log\left(\frac{1}{\tau}\right)}} + 1 \right\rfloor \leq Number\_of$$

$$Clusters \leq \left\lceil \frac{W^2}{2\left(\theta_1^2 \sqrt[\theta_2]{\left(\log\left(\frac{1}{\tau}\right)\right)^2}\right)} + \frac{W}{\theta_1 \sqrt[\theta_2]{\log\left(\frac{1}{\tau}\right)}} + 1 \right\rceil$$

(13)

The *equation no.* (13) shows the relation between the number of clusters and the threshold used for data correlation. If the threshold increases, the number of clusters with in the sensor field will get increases and vice versa. Thus we should choose appropriate threshold for clusters to perform data correlation in the spatial domain. Since the data are spatially correlated among the sensor nodes, there exist overlapping of clusters in the sensor region Z as shown in Figure 1. *Equations no* (12) and (13) derives how the clusters are overlapped among them in the sensor region Z. Hence it is important to find a distributed algorithm for clusters that can separate out the clusters from each other in the sensor region. Overlapping of cluster can sense the same correlated data among the sensor nodes and send the overlapped data to the sink node. It is like utilizing the same resource among the sensor nodes .Hence it leads to wastage of energy resource among the clusters and increases the data redundancy. Here we propose a distributed algorithm for cluster to overcome this problem for spatially correlated data and form non-overlapped irregular clusters in the sensor region Z.

___

***Algorithm I: Distributed clustering algorithm for spatially correlated data in sensor field Z.***
___

- Let U be the set of cluster head (CH) nodes deterministically deployed uniformly in sensor region Z.
- Let V be the set of sensor (normal) nodes randomly deployed in sensor region Z.
- Let d(a,b) be the Euclidian distance between node a and b.
- Let $d_v$ be the distance from node v to the nearest CH node.
- Initialize $d_v = \infty$
- Initialize $CH_v = 0$
- for $v \in V$
-     for $u \in U$
-         if $d(v,u) < d_v$
-             $d_v = d(v,u)$
-             $CH_v = u$
-         endif
-     endfor
- endfor
___

Thus the *Algorithm-I* finds the distance between shortest path from the normal nodes to each *CH* nodes where the data are spatially correlated among them and find out the non overlapping irregular distributed cluster in the sensor field Z. Thus the initial input is the overlapping clusters and the final output is the non overlapping irregular clusters formation in the spatial domain. In the next section we will perform the data accuracy for each individual distributed cluster.

### C. Data Accuracy Model for Distributed Clusters

We get distributed non overlapping irregular clusters in the sensor region Z as explained in the previous section. Each cluster has different set of sensor nodes. Each cluster can perform data accuracy at the *CH* node before data aggregation. For our convenience, we choose a single cluster of *M* set of sensor nodes to perform the data accuracy for cluster where $||M||=m$ are number of sensor nodes in the cluster. *M* set of sensor nodes which form a cluster can sense the tracing point and check the data accuracy at the *CH* node as shown in Figure 2. The data accuracy for *M* set of sensor nodes which form a cluster are verified before data aggregation process at the *CH* node for each cluster. The data accuracy is performed to check the data received at *CH* node for each cluster are accurate and doesn't contain any redundant data. They may reduce the communication overhead. For the simplicity of our model, Single tracing point is sensed and measured by only one single cluster. Hence in the sensor region Z, each distributed cluster can sense and measure a single tracing point.

In our model, there is a single tracing point *S* which can sense and measure by a respective cluster of *M* set of sensor nodes. Notation used in data accuracy for distributed clusters:

S = tracing point

$\hat{S}$ = estimation of tracing point

$S_i$ = physical phenomenon of *S* sensed by node *i* with no noise

$\hat{S}_i$ = estimation of $S_i$

$S_{CH}$ = physical phenomenon of $S$ sensed by cluster head node with no noise

$\hat{S}_{CH}$ = estimation of $S_{CH}$

$X_i$ = observed sample of $S_i$ by node i

$Y_i$ = observed sample of $X_i$ under transmission noise

$Z_i$ = observed sample of $Y_i$ under power constraint

$N_i$ = noise under additive white Gaussian noise (AWGN)

$N_{t_i}$ = transmission noise under AWGN

$\|M\|=m$ = total number of sensor node in a cluster

$d_{S,i}$ = distance between S and node i

$d_{S,CH}$ = distance between S and CH node

$d_{CH,i}$ = distance between CH node and node i

$d_{i,j}$ = distance between nodes i and j

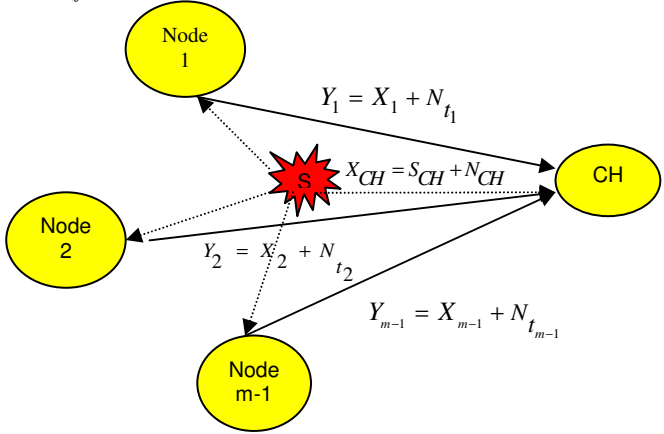

Figure 2: Data accuracy model for distributed cluster

Here we consider the mathematical analysis for data accuracy for the single cluster in the sensor region Z. Thus every cluster distributed in the sensor region can verify its data accuracy before data aggregation at the *CH* node. Once this procedure is being done by *CH* nodes for all the distributed clusters, they send the data to the sink node. Each sensor node *i* in the distributed cluster *M* can measure and observe the physically sensed data $S_i$ for tracing point S with observation noise $N_i$. Hence the observation and measurement made by the sensor node *i* in a given cluster is given by

$$X_i = S_i + N_i \quad \text{where } i \in M \quad (14)$$

The sensor node *i* can sense and measure the observe sample $X_i$ and transmits $X_i$ to cluster head node sharing wireless additive white Gaussian noise (AWGN) channel [9,14]. Hence the observation and measurement received by the CH node from other sensor nodes in the cluster with transmission noise $N_{t_i}$ over the AWGN channel is given by

$$Y_i = X_i + N_{t_i} = S_i + N_i + N_{t_i} \quad \text{Where } i \in M \text{ and } i \notin CH \quad (15)$$

We adopt uncoded transmission with finite number of sensor nodes for optimal point-to-point transmission [10] and consider the encoding power constraint value *P*, the measured value received by the *CH* are given by

$$Z_i = \sqrt{\frac{P}{(\sigma_{S_i}^2 + \sigma_{N_i}^2 + \sigma_{N_{t_i}}^2)}} Y_i = \alpha(S_i + N_i + N_{t_i}) \quad (16)$$

$$\text{where } i \in M \text{ and } i \notin CH$$

and $\quad \alpha = \sqrt{\dfrac{P}{(\sigma_{S_i}^2 + \sigma_{N_i}^2 + \sigma_{N_{t_i}}^2)}}$

*CH* node can sense and measure the tracing point *S* by finding the estimate of each physical phenomenon $S_i$ for node *i*. We take minimum mean square estimation (MMSE) for optimal decoding phenomenon [15] for uncoded transmission. *CH* node can find the MMSE for sensing and measuring the physical phenomenon $S_i$ extracted by sensor node *i* with observed sample $Z_i$ represented as

$$\hat{S}_i = \frac{E[S_i Z_i]}{E[Z_i^2]} Z_i \quad \text{where } i \in M \text{ and } i \neq CH \quad (17)$$

Since the sensor node *i* can sense and measure the physical phenomenon $S_i$ of *S*, we take independent identically distributed (i.i.d) Gaussian random variable with zero mean and variance $\sigma_S^2$ i.e $E[S]=0$, $\text{var}[S]=\sigma_S^2$ for tracing points. Similarly for sensing and measuring phenomenon of $S_i$, we assume $E[S_i]=0$, $\text{var}[S_i]=\sigma_{S_i}^2$. We also have taken the observation noise $N_i$ and transmission noise $N_{t_i}$ with an independent identically distributed Gaussian random variable with variances $\sigma_{N_i}^2, \sigma_{N_{t_i}}^2$ respectively with zero means.

Hence $E[N_i]=0$, $E[N_{t_i}]=0$, $\text{var}[N_i]=\sigma_{N_i}^2$, $\text{var}[N_{t_i}]=\sigma_{N_{t_i}}^2$ respectively.

Thus, $\quad E[S_i Z_i] = \alpha \sigma_{S_i}^2$

$$E[Z_i^2] = \alpha^2 (\sigma_{S_i}^2 + \sigma_{N_i}^2 + \sigma_{N_{t_i}}^2)$$

Thus the estimation of $\hat{S}_i$ is given by

$$\hat{S}_i = \frac{\sigma_{S_i}^2}{(\sigma_{S_i}^2 + \sigma_{N_i}^2 + \sigma_{N_{t_i}}^2)} (S_i + N_i + N_{t_i}) \quad (18)$$

where $i \in M$ and $i \notin CH$

$$\beta_i = \frac{\sigma_{S_i}^2}{(\sigma_{S_i}^2 + \sigma_{N_i}^2 + \sigma_{N_{t_i}}^2)} \quad \text{for } 0 < \beta_i < 1 \quad (19)$$

The *CH* node can also sense and measure the physical phenomenon of data independent of all the sensor nodes with out any transmission noise. Hence *CH* node can measure the physical phenomenon $S_{CH}$ of tracing points *S* and doesn't require the uncoded transmission for optimal decoding scheme with power constraint *P*, it can calculate the MMSE for physical phenomenon $S_{CH}$ from the observation $X_{CH}$ (where $X_{CH}=S_{CH} + N_{CH}$) represented as

$$\hat{S}_{CH} = \frac{E[S_{CH} X_{CH}]}{E[X_{CH}^2]} X_{CH} \quad (20)$$

The observation noise $N_{CH}$ for *CH* node can be given as i.i.d Gaussian random variable with zero mean and variance $\sigma_{N_{CH}}^2$ we get

$$E[S_{CH} X_{CH}] = \sigma_{S_{CH}}^2$$

$$E[X_{CH}^2] = (\sigma_{S_{CH}}^2 + \sigma_{N_{CH}}^2)$$

Thus the estimation of $\hat{S}_{CH}$ is given by

$$\hat{S}_{CH} = \frac{\sigma_{S_{CH}}^2}{(\sigma_{S_{CH}}^2 + \sigma_{N_{CH}}^2)} (S_{CH} + N_{CH}) \quad (21)$$

Where $\beta_{CH} = \frac{\sigma_{S_{CH}}^2}{(\sigma_{S_{CH}}^2 + \sigma_{N_{CH}}^2)} \quad \text{for } 0 < \beta_{CH} < 1 \quad (22)$

Hence we get two constraint factors $\beta_i$ and $\beta_{CH}$ from *equations no (19) and (22)*, which perform data accuracy under Gaussian noise for each distributed cluster. Therefore *M* set of sensor nodes forms a cluster and control the sensing and measuring phenomenon of measurement for moisture content in a soil. We measure the data accuracy performed by each distributed cluster at the *CH* node in the sensor region. To find the estimate of tracing point *S* done by the cluster at the *CH* node, we compute the average of the entire MMSE observation sample done by *m* sensor nodes and the expression for average estimate is given by

$$\hat{S}(M) = \frac{1}{m} \left[ \sum_{i=1}^{m-1} \beta_i (S_i + N_i + N_{t_i}) + \beta_{CH}(S_{CH} + N_{CH}) \right] \quad (23)$$

The data accuracy $D(M)$ for the estimation of every distributed cluster with different set of sensor nodes in the sensor region is defined in terms of the expectation of the error between the actual value of tracing point and the mean square average estimates value of *M* set of sensor nodes in the cluster. Hence we adopt mean square error between *S* and $\hat{S}(M)$ to find data accuracy estimation for every cluster in the sensor region is given by

$$D(M) = E[(S - \hat{S}(M))^2]$$

$$D(M) = E[S^2] - 2E[S\hat{S}(M)] + E[\hat{S}(M)^2] \quad (24)$$

The normalized [2] data accuracy $D_A(M)$ for each distributed cluster in the sensor region is given as

$$D_A(M) = 1 - \frac{D(M)}{E[S^2]}$$

$$D_A(M) = \frac{1}{E[S^2]} [2E[S\hat{S}(M)] - E[\hat{S}(M)^2]] \quad (25)$$

Thus every distributed cluster can perform the normalized data accuracy at their *CH* node before data aggregation and finally send the data to the sink node. The normalized data accuracy for each cluster in the sensor region can be implemented in spatial correlation model explained in the next section.

### D. Distributed Cluster-based Spatially Correlated Data Accuracy Model

In this section, a spatial correlation model is framed for normalized data accuracy for each cluster in the sensor region. Since distributed clusters are formed in the sensor region, every cluster can estimate its data accuracy to sense and measure their representative tracing point. Each tracing point is sensed and measured by a single cluster of sensor nodes and finally determines the data accuracy for that cluster in the sensor region. Here we derive a mathematical model for the distributed cluster where all the sensed data are spatially correlated among them. These spatial correlations among data are achieved by *M* set of sensor nodes. We model spatially correlated physical phenomenon of sensed data as joint Gaussian random variables (JGRV's) [5] as follows:

Step 1: $E[S] = 0$, $E[S_i] = 0$, $E[S_{CH}] = 0$;
$E[N_i] = 0$, $E[N_{t_i}] = 0$, $E[N_{CH}] = 0$

Step 2: $var[S] = \sigma_S^2$, $var[S_i] = \sigma_{S_i}^2$, $var[S_{CH}] = \sigma_{S_{CH}}^2$;
$var[N_i] = \sigma_{N_i}^2$, $var[N_{t_i}] = \sigma_{N_{t_i}}^2$, $var[N_{CH}] = \sigma_{N_{CH}}^2$

Step 3: $cov[S, S_i] = \sigma_S^2 corr[S, S_i]$
$cov[S, S_{CH}] = \sigma_S^2 corr[S, S_{CH}]$
$cov[S_i, S_j] = \sigma_S^2 corr[S_i, S_j]$
$cov[S_{CH}, S_i] = \sigma_S^2 corr[S_{CH}, S_i]$

*Step 4:* $E[S, S_i] = \sigma_S^2 corr[S, S_i] = \sigma_S^2 \rho(s,i) = \sigma_S^2 K_V(d_{s,i})$

$E[S, S_{CH}] = \sigma_S^2 corr[S, S_{CH}] = \sigma_S^2 \rho(s,CH) = \sigma_S^2 K_V(d_{S,CH})$

$E[S_i, S_j] = \sigma_S^2 corr[S_i, S_j] = \sigma_S^2 \rho(i,j) = \sigma_S^2 K_V(d_{i,j})$

$E[S_{CH}, S_i] = \sigma_S^2 corr[S_{CH}, S_i] = \sigma_S^2 \rho(CH,i) = \sigma_S^2 K_V(d_{CH,i})$

Explanation of step *1-2* is already given in the *section II(C)*. Using step *3-4,* we demonstrate the covariance model [16] for spatially correlated data for each distributed cluster in the sensor region Z. To clarify the covariance model say $cov[S_i, S_j] = E[S_i, S_j] = \sigma_S^2 corr[S_i, S_j] = \sigma_S^2 \rho(i,j) = \sigma_S^2 K_V(d_{i,j})$ where $d_{ij} = \| S_i - S_j \|$ represents the Euclidean distance between node $n_i$ and $n_j$ and $K_V(.)$ is the correlation model for spatially correlated data in a single cluster. The covariance function is non-negative and decrease monotonically with the Euclidean distance $d_{ij} = \| S_i - S_j \|$ with limiting values of 1 at *d=0* and of 0 at $d = \infty$. We have taken power exponential model [17] i.e. $K_V^{P.E}(d_{i,j}) = e^{-(d_{i,j}/\theta_1)^{\theta_2}}$, $\theta_1 > 0; \theta_2 \in (0,2]$ $\theta_1$ is the '*Range parameter*' and $\theta_2$ is the '*Smoothness parameter*'.

Using *(15)* and *(23)* in *(25)*, we perform the normalized data accuracy with spatial correlation model for every distributed cluster in the sensor region given as follows:

$$D_A(M) = \frac{1}{m}\left[\beta_i(2\sum_{i=1}^{M-1} e^{(-d_{(S,i)}/\theta_1)^{\theta_2}} - 1) + 2\beta_{CH} e^{(-d_{(S,CH)}/\theta_1)^{\theta_2}}\right]$$

$$-\frac{1}{m^2}\left[\beta_i(\beta_i \sum_{i=1}^{M-1}\sum_{j\neq i}^{M-1} e^{(-d_{(i,j)}/\theta_1)^{\theta_2}} - 1) + \beta_{CH}(2\beta_i \sum_{i=1}^{M-1} e^{(-d_{(CH,i)}/\theta_1)^{\theta_2}} + \beta_{CH})\right]$$

(26)

The *equation no.* (26) shows that the normalized data accuracy $D_A(M)$ for each cluster depends upon *m* sensor nodes and factors $\beta_i$ and $\beta_{CH}$ respectively. Since we get a normalized data accuracy at each *CH* node for each cluster, we construct a spatial correlation model given by *equation no. (26)* for each individual distributed cluster in the sensor region. The spatial correlation model for each distributed cluster can be explained as follows:
> Each sensor node *i* can sense a tracing point *S* in each distributed cluster where $i \in M$ and $i \notin CH$ node
> *CH* node itself can sense the tracing point S in each distributed cluster.
> A spatial correlation between node *i, j* in each distributed cluster where $i,j \neq CH$ node.
> Each sensor node *i* transmits the sensed data to the *CH* node in each distributed cluster where $i \in M$ and $i \notin CH$.

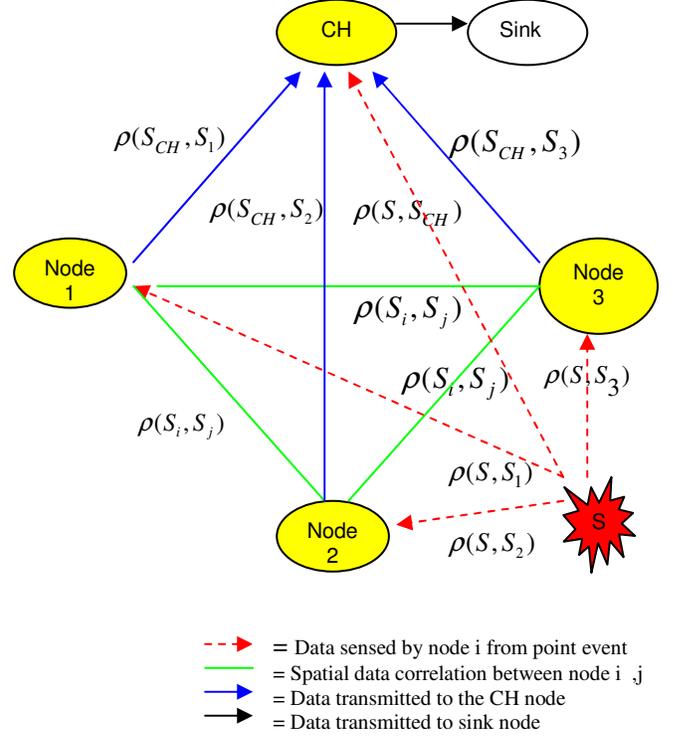

Figure 3: Spatial correlation model for distributed cluster

Thus each distributed cluster formed in the sensor region has different set of sensor nodes. Hence each cluster can perform the normalized data accuracy at the *CH* node before data aggregation. The purpose of verifying the data accuracy for each cluster is to confirm that the most accurate data send by *m* set of sensor nodes can aggregate at the *CH* node rather than aggregating all the redundant data at the *CH* node. To visualize the correlation model for distributed cluster, we take an example where *m=4* sensor nodes and out of *m* sensor nodes one node is chosen as a *CH* node as shown in Figure 3. Once we estimate the data accuracy at the *CH* node for each distributed cluster, the most accurate data get aggregated and finally send to the sink node.

### III. SIMULATION RESULTS

In the first simulation setup, twenty five *CH* nodes are deterministically deployed uniformly and hundred sensor (normal) nodes are deployed randomly in a wireless sensor field of *120 m X 120 m* based sensor topology as shown in Figure 1 . Each *CH* node performs the data accuracy for their respective cluster. Hence each cluster can sense and measure a single tracing point randomly located in each cluster region. Once each cluster can sense and measure their respective tracing point, it performs the data accuracy at *CH* node and finally transmits the data to the sink node.

| #CH Nodes | Associated Nodes (Normal Nodes) | Data Accuracy |
|---|---|---|
| CH1 | 2  21  59  78 | 0.837847 |
| CH2 | 1  6  7  8  13  35  43  76  92  93 | 0.843960 |
| CH3 | 11  17  22  69  84  98 | 0.866797 |
| CH4 | 10  46  62 | 0.694458 |
| CH5 | 4  40  58  87 | 0.833017 |
| CH6 | 15  25  32  33  53  73  81 | 0.820673 |
| CH7 | 36  41  57  61  74  80  83 | 0.862045 |
| CH8 | 13  19  28  31  49  85  95 | 0.793657 |
| CH9 | 9  29  37  38 | 0.882088 |
| CH10 | 20  23  63  75  77  79  88  91  97 | 0.857425 |
| CH11 | 44  51  66  86  99 | 0.820772 |
| CH12 | 45  50  55  89 | 0.809979 |
| CH13 | 5  18  24  47  48  52  82 | 0.813055 |
| CH14 | 27  30  34  39  71  100 | 0.756650 |
| CH15 | 26  60  72 | 0.787127 |
| CH16 | 70 | 0.714302 |
| CH17 | 65  94 | 0.854421 |
| CH18 | 68  90 | 0.873163 |
| CH19 | 42 | 0.705224 |
| CH20 | 56 | 0.759352 |
| CH21 | 67 | 0.730805 |
| CH22 | 12 | 0.799681 |
| CH23 | 96 | 0.739846 |
| CH24 | 14  16 | 0.894157 |
| CH25 | 54  64 | 0.843685 |

*Table 1: Data Accuracy for each distributed cluster*

According to our proposed *algorithm-I* discuss previously, each *CH* node can form the cluster with their associated sensor nodes. Once the sensor nodes are associated with each *CH* node, they form distributed clusters in the sensor region Z. Thus twenty five *CH* nodes can form twenty five individual non-overlapping distributed clusters. Each distributed cluster can perform the data accuracy at their respective *CH* node as shown in *Table-1*. Similarly in the second simulation set up as shown in *Table-2,* we perform hundred runs for each *CH* nodes associated with their respective sensor nodes and find their average data accuracy for each cluster.

| #CH Nodes | Average Data Accuracy | #CH Nodes | Average Data Accuracy |
|---|---|---|---|
| CH1 | 0.8494 | CH14 | 0.7327 |
| CH2 | 0.8731 | CH15 | 0.7778 |
| CH3 | 0.8765 | CH16 | 0.9662 |
| CH4 | 0.8734 | CH17 | 0.8001 |
| CH5 | 0.8468 | CH18 | 0.7706 |
| CH6 | 0.8401 | CH19 | 0.9662 |
| CH7 | 0.8364 | CH20 | 0.8111 |
| CH8 | 0.7975 | CH21 | 0.8135 |
| CH9 | 0.9033 | CH22 | 0.9736 |
| CH10 | 0.8615 | CH23 | 0.9047 |
| CH11 | 0.7942 | CH24 | 0.8343 |
| CH12 | 0.8171 | CH25 | 0.8352 |
| CH13 | 0.7796 | | |

*Table 2: Average Data Accuracy for each distributed cluster*

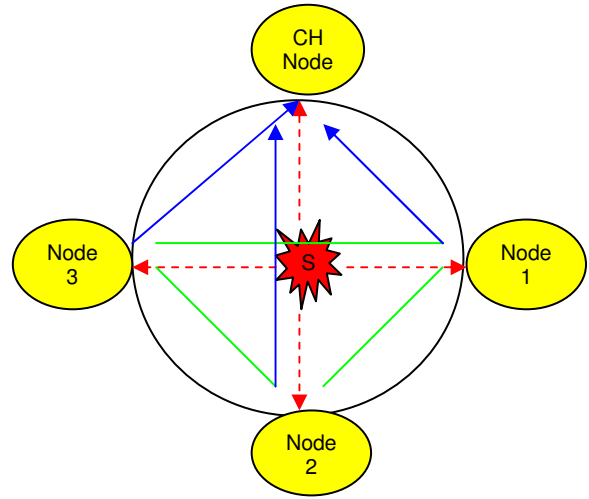

Figure 4: Deployed sensor nodes in circular cluster topology

In the third simulation set up, we take a single circular cluster of *m=4* sensor nodes which can sense and measure a tracing point. We put *m* sensor nodes in a deployed circular cluster and a tracing point *S* located at the centre of the deployed circular cluster. i.e $d_{S,i}$ *(where i=1,2,3)* and $d_{S,CH}$ are equidistance as shown in the Figure 4. Here we have fixed the number of *m* sensor nodes and vary the distance from the tracing point *S* to *m* sensor nodes. As we increase the radius of the deployed circular cluster for $d_{S,i}$ and $d_{S,CH}$ with same proportion, $D_A(M)$ decreases i.e. the distance from the tracing point $S$ to the $m$ sensor nodes increases as shown in Figure 5. We put $\theta_1 = \{50,100\}$ and $\theta_2 = 1$ for our statistical data performance for the normalized data accuracy $D_A(M)$.

field). We verified that $D_A(M = 4)$ is 0.6333 when $\theta_1 = 50$ as shown in Figure 8. If we increase $\theta_1 = 400$, then $D_A(m = 4) = 0.911$.

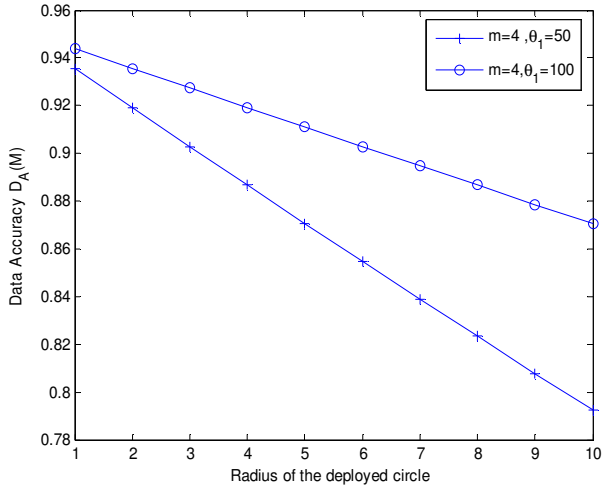

Figure 5: Data accuracy versus radius of the circular cluster

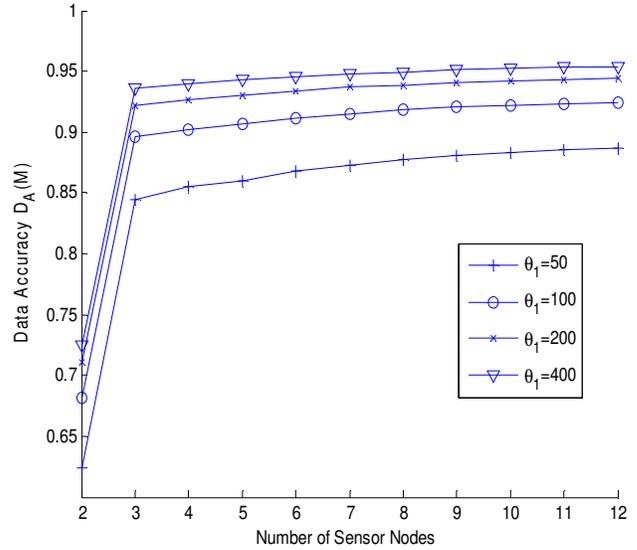

Figure 6: Data accuracy versus number of sensor nodes in a cluster

In the fourth simulation setup, the distance from the tracing point $S$ to $m$ sensor nodes is fixed in the deployed circular cluster of radius =5 metre. We increase the number of sensor nodes with a fixed distance from the tracing point $S$ i.e we increase $m$ sensor nodes with fixed deployed circular cluster of radius 5 metre. At first, we put $m=2$ (one CH node and one sensor node) which shows that the data accuracy is very poor with its value in between 0.6 to 0.75 for $\theta_1 = \{50,100,200,400\}$. The reason is that there is only one sensor node which shows that the third condition of spatial correlation model given in *section II(D)* doesn't satisfies the $D_A(M)$ at the CH node. But if we put m=3 (one cluster head and two sensor nodes), there is a drastic improvement of $D_A(M)$ since all the conditions for spatial correlation model are satisfied. The Figure-6 also shows that five to eight nodes are sufficient to perform the $D_A(M)$ for the cluster, if the distance from tracing point to $m$ sensor nodes with deployed circular cluster of radius is 5 metre.

For the simplicity of our model, we perform the fifth simulation set up where we have simulated a wireless sensor field (900 metre$^2$) of 5m X 5m grid based single cluster topology with a fixed tracing point *(S)* at the centre and a CH node on the corner edge with 47 sensor nodes distributed uniformly in the grid based cluster topology as shown in Figure 7. Our assumptions is that cluster of $m$ sensor nodes are in the sensing range of the tracing point *(S)*. Initially we put $m=4$ (one cluster head node and three sensor nodes located at the four extreme corner of sensor

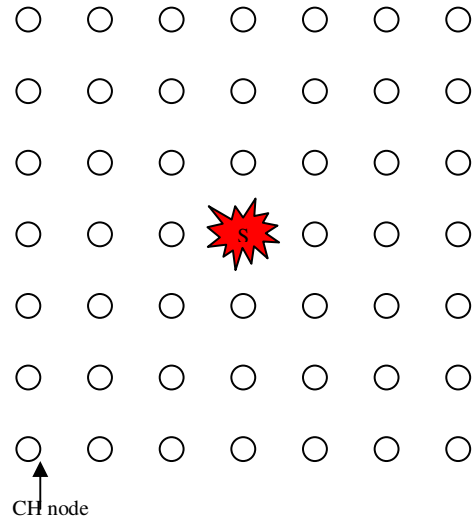

Figure 7: Sensor nodes deployed in grid topology

This shows that $\theta_1$ control as how fast the spatially correlated data decays with distance between sensor nodes and the tracing point. Hence it is always suitable to take the value of $\theta_1$ large for large sensor field to get $D_A(M)$ in an efficient way. Now we increase cluster of $m$ sensor nodes with increment of four sensor nodes every time concentrating towards tracing point till $m$ sensor nodes are able to sense and measure the tracing point $S$ in the region. As we increase the sensor nodes, the data accuracy $D_A(M)$ also get increases. Hence for 900 metre$^2$ sensor field, 15 to 20 sensor nodes are sufficient to give $D_A(M)$ of 0.944 for $\theta_1$ =400 and $D_A(M)$ remains approximately constant still we increase the number of sensor nodes for the cluster. We plot in the Figure-8 for the $D_A(M)$ versus node density for a cluster. Node density is defined as the number of sensor nodes per unit area in a single cluster. Hence it is needless to choose so many sensor nodes to achieve data accuracy for the cluster in sensor field to sense and measure a tracing point.

the $D_A(M)$ remains approximately constant for the cluster. If we continuously increase the value of $\theta_1$ the average $D_A(M)$ remains approximately constant since it achieve the saturation level in the cluster. Finally the output graph shows distortion in the signal due to additive white Gaussian noise components.

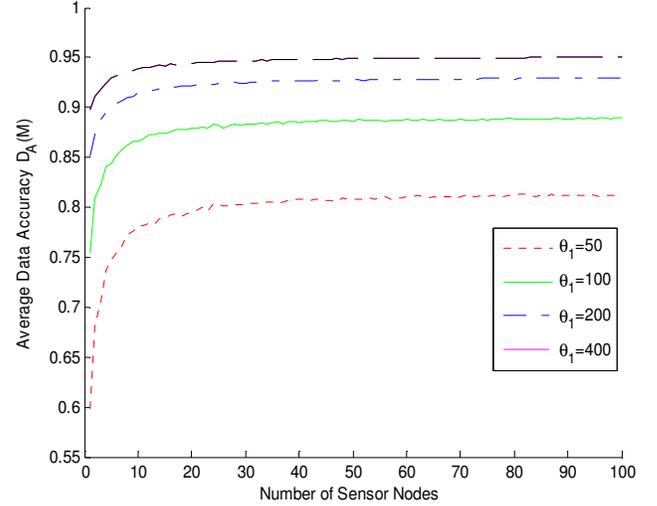

*Figure 9: Average data accuracy versus number of sensor nodes in a single cluster*

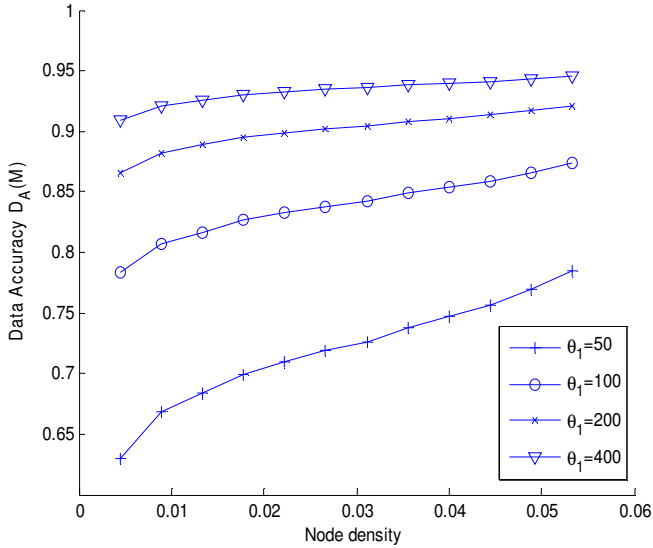

*Figure 8: Data accuracy vs. node density in a single cluster*

In the sixth simulation setup, we take a single cluster of *m* sensor nodes randomly deployed in a region (30 X 30 = 900 metre$^2$) that sense and measure a tracing point. We fix the tracing point at x,y (15,15) coordinate and *CH* node at x,y (0,0) coordinate with 99 sensor nodes randomly deployed in the region. For each run we verify $D_A(M)$ with respect to randomly deployed cluster of *m* sensor nodes. Finally we verify for 100 runs and find the average $D_A(M)$ for the cluster of *m* sensor nodes. Figure 9 shows that if the value of $\theta_1$ =400, $D_A(M)$ is 0.944 for 10 to 15 sensor nodes. If we continuously increase the number of sensor nodes the $D_A(M)$ remains approximately same. Hence it is useless to deploy sensor nodes beyond 15 sensor nodes because 10 to 15 sensor nodes are sufficient to give approximately the same $D_A(M)$ for the cluster with $\theta_1$ =400. Again if we constantly increases $\theta_1$, average $D_A(M)$ also get increases for the cluster of *m* sensor nodes. But after certain approximate value of $\theta_1$

Since the data are spatially correlated in the sensor region, we propose a distributed algorithm with non overlapping irregular cluster for the spatially correlated data in the sensor region. Each distributed cluster can perform $D_A(M)$ before data aggregation at their respective *CH* node. Hence it is important to sense and measure the most appropriate (accurate) data send by each distributed cluster at the *CH* node rather than aggregating all the redundant data at their respective *CH* node. Thus it can reduce the data redundancy. Since the data accuracy is performed by each distributed cluster, we verified from the simulation results that there exists a minimal set of sensor nodes with optimal cluster which is sufficient to give approximately the same $D_A(M)$ as achieved by the each distributed cluster. Therefore the time complexity done at each *CH* node of respective distributed cluster for aggregating the most accurate data send by their respective optimal cluster will be less. Thus we find an optimal cluster from each distributed cluster which can reduce the data redundancy and communication overhead.

In the fifth simulation setup, a grid based single cluster is formed where we deployed m=48 sensor nodes uniformly. We examine that 15 to 20 nodes are sufficient to perform $D_A(M)$ =0.944 for $\theta_1$ =400 in 900 metre$^2$ cluster region. Similarly in sixth simulation setup a cluster with m=100 sensor nodes are randomly deployed in 900 metre$^2$ region and we get 10 to 15 sensor nodes are sufficient to perform

$D_A(M) = 0.944$ for $\theta_1 = 400$. Therefore it is unnecessary to choose so many sensor nodes in 900 metre² region as $D_A(M)$ remains approximately same as it achieve the saturation level still we increase *m* sensor nodes in the cluster. Hence we have *P* minimal set of sensor nodes with optimal cluster which is sufficient to give approximately the same $D_A(M)$ by *M* set of sensor nodes in each distributed cluster as shown by Venn diagram in Figure 10.

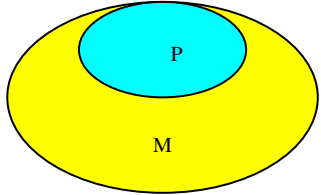

*Figure 10: Venn diagram for optimal cluster in each distributed cluster*

## IV. CONCLUSIONS

In this paper we investigate that the data are spatially correlated among sensor nodes and form clusters in the sensor region. Since the data are highly correlated in the spatial domain, the sensor nodes form regular overlapping clusters among them in the sensor region. Overlapping of cluster can sense and measure the same correlated data among the clusters. Thus to overcome this situation, we constructed a distributed clustering algorithm with data accuracy model .We perform data accuracy for each distributed cluster. We find that the most accurate data send by the distributed cluster can aggregate at the *CH* node rather than aggregating all the redundant data at their respective *CH* node. We demonstrate by simulation that the data accuracy for a single cluster depend on number of sensor nodes and their exist an optimal cluster which is adequate to sense and measure the tracing point to perform approximately the same data accuracy level achieve by single cluster. Finally we conclude that the data accuracy performed for each distributed cluster can reduce the data redundancy and communication overhead.

## REFERENCES

[1] I.F Akyuildz ,W.Su , Y. Sankarasubramanian and E. Cayirci, "A survey on sensor Networks ",IEEE Communcations Magazine, vol .40 , pp.102-114 ,Aug 2002.

[2] S.S. Pradhan , K. Ramchandran ,"Distributed Source Coding : Symmetric rates and applications to sensor networks", in procecding of the data compressions conference 2000,pp.363-372.

[3] A. Abbasi and M. Younis , "A survey on clustering algoirthms for wireless sensor networks " Computer communications , vol-30 , n-14-15 ,pp-2826-2841, 2007.

[4] W.B Heinzelman , Anantha P. Chandrakasan, "An Application Specific Protocl Architecture for Wireless Microsensor Networks " , IEEE transactions on wireless communications , vol. no 4 , Oct-2002.

[5] Georgios Smaragdakis, Ibrahim Matta ,Azer Bestavros , " SEP : A stable Electon Prtocol for cluster hetrerogenous wireless sensor networks "

[6] Chongqing Zhang , Binguo wang , Shen Fang , Zhe Li , " Clustering Algorithms for wireless sensor networks using spatial data correlation ", International conference on information and Automation , pp-53-58 .june 2008.

[7] Zhikui chen , Song Yang , Liang Li and Zhijiang Xie , " A clustering Approximation Mechinism based on Data Spatial Correlation in Wireless sensor Networks ", Proceedings of the 9[th] international conferenses on wireless telecommunication symposium -2010.

[8] Ali Dabirmoghaddam , Majid Ghaderi , Carey Williamson , " Energy Efficient Clustering in wireless Sensor Networks with spatially correlated dara " IEEE infocom 2010 proceedings.

[9] Kang Cai, Gang Wei and Huifang Li,"Information Accuracy versus Jointly Sensing Nodes in Wireless Sensor Networks" IEEE Asia Pacific conference on curcuit and systems 2008 ,pp.1050-1053.

[10] M.Gastpar, M. Vetterli, " Source Channel Communication in Sensor Networks ", Second International Workshop on Information Processing in Sensor Networks (IPSN'2003).

[11] Varun M.C,Akan O.B and I.F Akyildiz, " Spatio-Temporal Correlation : Theory and Applications Wireless Sensor Networks" , Computer Network Journal (Elsevier Science ), vol. 45 , pp.245-259 , june 2004.

[12] Jyotirmoy karjee , H.S Jamadagni , "Data Accuracy Estimation for Cluster with Spatially Correlatd Data in Wireless Sensor Networks " ,to be published in the proccedings ICISCI-2011, Harbin ,China

[13] Huifang Li, Shengming Jiang ,Gang Wei ,"Information Accuracy Aware Jointly Sensing Nodes Selection in Wireless Sensor Networks ",MSN 2006 , LNCS 4325 , pp.736-747.

[14] T.J. Goblick ," Theoritical Limitions on the transmission of data from analong sources",IEEE Transaction Theory , IT-11 (4) pp.558-567 ,1965.

[15] V.Poor ," An Introduction to Signal Detection and Estimation ",Second edition , Springer ,Berlin 1994.

[16] J.O. Berger , V.de Oliviera and B.Sanso ," Objective Bayesian Anylysis of Spatially correlated data ".J.Am.Statist. Assoc. Vol-96,pp.1361-1374,2001.

[17] De Oliveira V, Kedan B and Short D.A , " Bayesian predication of transformed Gaussian random fields" Journal of American statistical Association 92, pp.1422-1433.

[18] L.Guo , F chen , Z Dai , Z. Liu ,,"Wireless sensor network cluster head selection algorithm based on neural networks" , PP-258-260 , International conference on Machine vision and human machine interference, 2010.

[19] T.Minming , N Jieru , W Hu , Liu Xiaowen " A data aggregation Model for underground wireless sensor network" Vol-1, pp-344-348 , WRI world congress on computer science and information engineering, 2009 .

[20] Jyotirmoy karjee , Sudipto Banerjee , " Tracing the Abnormal Behavior of Malicious Nodes in MANET ", Fourth International conference on wireless communications , networking and Mobile Computing ,pp-1-7 Dalian-china -2008 .

[21] C.Y. cho , C.L Lin , Y.H Hsiao , J S wang , K.C yong " Data aggegation with spatially correlated grouping Techninques on cluster based WSNs" , SENSORCOMM ,pp-584-589, venice- 2010.

[22] Shirshu Varma , Uma shankar tiwary , " Data Aggregation in Cluster based wireless sensor Networks "Proceedings of the first International confernce on Intelligent human computer interaction , page-391-400 , part-5 , 2009.

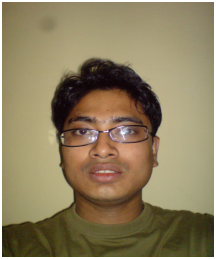
**Jyotirmoy Karjee** received his B.E (Electronics), M.E(Information Technology ) specialization in Network Security in 2003 and 2005 respectively. He worked in Prakriti Inbound Pvt. Ltd as a software engineer for a year and worked as a lecturer in Sikkim Manipal Institute of Technology , Sikkim till 2008. He is currently pursuring his Ph.D degree at Centre for Electronics Design and Technology , Indian Instutute of Science , Bangalore . His current research interests includes data accuracy estimation and data aggregation in wireless sensor networks.

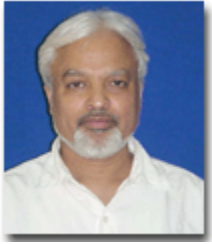
**Prof. H.S Jamadagni** received his M.E and Ph.D degree in Electrical & Communication Engineering from Indian Institute of Science ,Bangalore. Currently He is the professor at Centre for Electronics Design and Technology , Indian Institute of Science. He is one of the main coordinators for the intel higher education program and was the key mentors for various intel workshops in india. His current research work includes in the areas of embedded systems , VLSI for wireless networks and wireless sensor networks .